\documentstyle[aps,prl]{revtex}
\begin{document}

\draft

\title{Word Processors with Line-Wrap:  Cascading, Self-Organized Criticality,
Random Walks, Diffusion, Predictability}

\author{Wolfgang Bauer and Scott Pratt}

\address{Department of Physics and Astronomy and National Superconducting
Cyclotron Laboratory, Michigan State University, East Lansing, MI 48824-1321,
USA}

\date{June 28, 1995}

\maketitle

\begin{abstract}
We examine the line-wrap feature of text processors and show that adding
characters to previously formatted lines leads to the cascading of words
to subsequent lines and forms a state of self-organized criticality.  We
show the connection to one-dimensional random walks and diffusion problems,
and we examine the predictability of catastrophic cascades.
\end{abstract}

Composite systems may evolve to a critical state in which minor events may
trigger a chain reaction that can effect an arbitrarily large number of
constituents of the system.  This state was called {\em self-organized
criticality} \cite{BTW87,BC91,BC94}, and was first investigated for sandpiles.
Theoretically and experimentally \cite{GMH93}, avalanches in sandpiles show
power-law distributions characteristic of real earthquakes \cite{GR56,CL89}.
The absence of an intrinsic length scale is attributed to self-organized
criticality, where avalanches of all sizes contribute to keep the system
perpetually in a critical state.

So far, no analytic solution to the model of Ref.\ \cite{BTW87} has been
presented.  However, if one introduces a preferred direction, then an
exact analytic solution is possible \cite{DR89,D90}, and a connection to
the problem of two annihilating random walkers can be established.

Of particular interest is the question of predictability of catastrophic
avalanches.  In a recent experiment measuring the total mass of the sandpile
it was found that a running total of small avalanches can predict
the occurrence of large avalanches \cite{RVR94}.

Here we introduce what we believe to be the most simple (and most directly
connected to everyday experiences) example of self-organized criticality.
Consider a modern word processor with line-wrap feature and fixed maximum
number of characters per line.  Such a word processor
formats a paragraph without the explicit need to enter carriage returns or
line feed characters.  If a word is too long for a line, it is automatically
wrapped into the next line.  We consider an infinitely long paragraph
formatted by this word processor.  If one then adds another character to the
beginning of the first line, the last word of this line may be wrapped to
the second line, and a cascade of line-wraps may ensue.  If a steady stream
of characters or words is entered at the beginning of the first line, a
sequence of cascades on all scales results, and the line
lengths (excluding trailing blanks)
in the paragraph form a self-organized critical state.

Let us -- for the moment -- neglect temporal
correlations between the individual words shifted through the
paragraph.  (This assumes in practice that individual lines are
infinitely long -- an assumption we will relax again below.)  In addition, we
assume here for definiteness that the individual word lengths
including one trailing blank are evenly distributed between 2 and
$2 \langle\ell_w\rangle - 2$, where $\langle\ell_w\rangle$ is the average
word length.  (The probability for any given word length between these
two limiting values is then $(2 \langle\ell_w\rangle - 3)^{-1}$.)
For a fixed number of characters per line, the probability
to have $b_n$ trailing blanks in this line can be shown to be
\begin{equation}
  p(b_n) = \left\{
           \begin{array}{ll}
               \langle\ell_w\rangle^{-1} & \mbox{for $n=0,1$}\\
               \frac{2\langle\ell_w\rangle-2-b_n}
                    {2\langle\ell_w\rangle^2-3\langle\ell_w\rangle}
                                         & \mbox{for $n\leq 1$}
           \end{array}
           \right.
  \label{b_n}
\end{equation}
and consequently the average number of trailing blanks in a line is
\begin{equation}
  \langle b_n\rangle = \sum_{b_n=0}^{2\langle\ell_w\rangle-2} b_n\,p(b_n)
      = {\textstyle\frac{2}{3}} \langle\ell_w\rangle - 1 +
        {\textstyle\frac{1}{3}} \langle\ell_w\rangle^{-1}\ .
  \label{avb_n}
\end{equation}

\begin{figure}
  \vspace*{12cm}
  \includegraphics{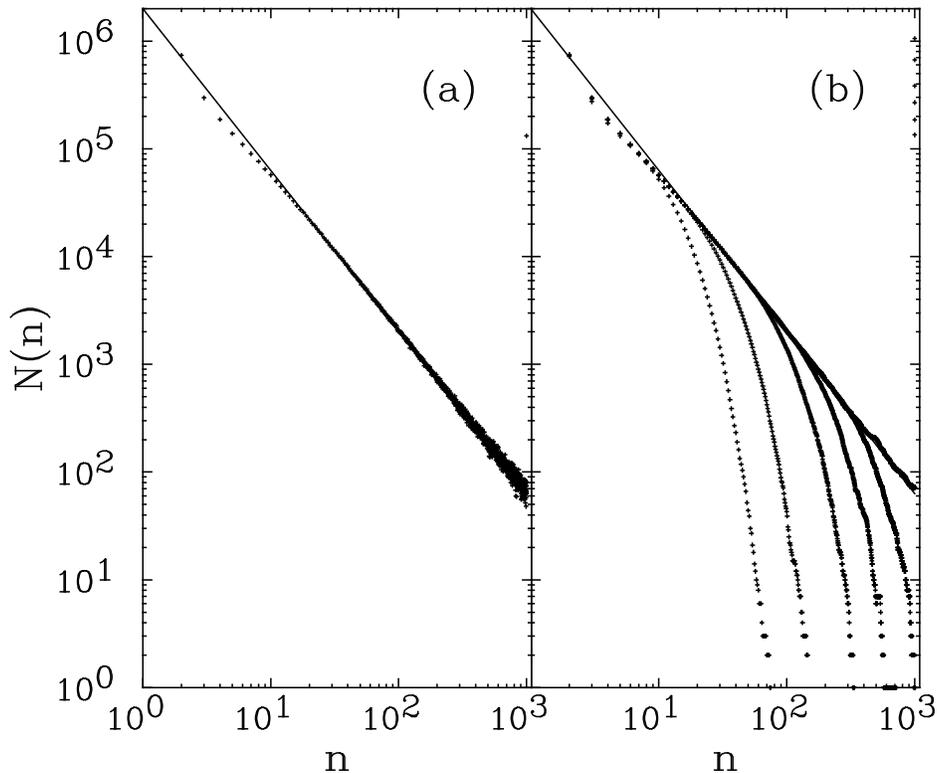}
  \caption[]{Cascade lengths distribution for the word processor with line-wrap
      feature.  (a) infinite line length, (b) finite line lengths,
      $\ell_l = 20$, 30, 50, 70, 100, 200, from left to right,
      respectively. The straight lines correspond to the asymptotic
      $n^{-3/2}$ solutions.}
  \label{Fig1}
\end{figure}

In Fig.\ \ref{Fig1} (a) we display (crosses)
the size distribution (= number of lines
affected) for cascades in this system, where the total number of lines
was chosen to be 10$^3$, and $\langle\ell_w\rangle = 6$.  A total of
$3\times 10^6$ words ($=1.8\times 10^7$ characters) were generated for this
plot.  One can clearly observe that for large number
of lines, $n$, the distribution approaches the power law limit (solid line),
\begin{equation}
  N(n) \propto n^{-3/2}\ .
  \label{power3/2}
\end{equation}
It may be of interest that some experimental manifestations of self-organized
criticality, such as the frequency distribution for power spectra in
superconducting vortex avalanches \cite{FWN95}, show similar power laws.

It is also of interest to compute the total activity, i.e.\ the total number
of characters moved to different lines.  The number distribution for this
activity also approaches a power law, with an exponent of approximately $-4/3$
very much reminiscent of the earthquake strength distribution found in
ref.\ \cite{GR56}.

We should point out here that none of our results depend on the type
of the word length distribution chosen.  We obtained for practical purposes
identical results with Poissonian word length distributions.

The result of Eq.\ (\ref{power3/2}) can be understood by formulating the
problem in terms of a random walk.  One step in this random walk
is the change in the number of blanks in a given line caused by a cascade
passing through.  To derive the step size distribution,
we realize that pushing a word of length
$\ell$ from line $n$ to $n+1$ increases the number of trailing blanks
in line $n$ by $\ell$.  Conversely, pushing a different word of length $\ell'$
from line $n-1$ into line $n$ decreases the number of trailing blanks in
line $n$ by $\ell'$.  The probability distribution for a
change $\Delta b_n$ in the number of trailing blanks in line
$n$ is then
\begin{equation}
  P(\Delta b_n) = \sum_{b_n=0}^{\infty} p(b_n)\,p(b_n\!-\!\Delta b_n)
  \label{Deltabn}
\end{equation}
where $p(b_n)$ is the probability distribution for trailing blanks as given by
Eq.\ (\ref{b_n}).  The probability distribution $P(\Delta b_n)$ is symmetric
about $\Delta b_n = 0$ and approximately triangular in shape.  Thus its mean
is 0, and its variance is finite.

The total number of
characters moved through line $n$ by the cascade is
\begin{equation}
  c_n = - \sum_{i=1}^{n-1} \Delta b_i\ .
\end{equation}
If for any $n$, we have $c_n \leq b_n$, then the cascade terminates.  Thus we
see that our cascading problem is equivalent to a random walk
problem with step size distribution given by Eq.\ (\ref{Deltabn}).
The result of Eq.\
(\ref{power3/2}) is the solution of the return-to-the-origin
problem for a one-dimensional random walk.

It is, perhaps, more instructive to consider the
corresponding
continuum diffusion problem.  The diffusion equation is
\begin{equation}
  \partial_t f(x,t) = D\, \partial^2_x f(x,t)
\end{equation}
with the boundary condition $f(0,t) = 0$ and the solution
\begin{equation}
  f(x,t) = \frac{x}{4\sqrt{\pi D^3 t^3}}\, \exp[-x^2/4Dt]\ ,
  \label{diffsol}
\end{equation}
where $t$ corresponds to the number of lines, $n$, in the random walk,
and $x$ is the distance of the random walk to 0.
$D$ is the diffusion constant and
can be calculated from the second moment of the random walk step size
distribution, Eq.\ (\ref{Deltabn}),
\begin{equation}
  D = \langle \Delta b_n^2\rangle / 2\ .
\end{equation}
($D\approx 6.3$ for the parameters used to produce Fig.\ \ref{Fig1}.)
For the current at the origin we obtain
\begin{equation}
  J(t) = D\,\partial_x f(x,t)|_{x=0} \propto t^{-3/2}\ ,
\end{equation}
in agreement with the numerical finding of Fig.\ \ref{Fig1}.

We now proceed to study the case where we
include all of the temporal correlations entailed by pushing an ordered
(but individually randomly selected) sequence of words through our word
processor.  This is the case for finite line lengths.
Inserting a number of characters equal to the
line length, $\ell_l$, will result in a completely new first line,
pushing the old first
line into the second, and so on.  Thus we get catastrophic cascades (= cascades
involving all lines -- 1000 in the specific example considered here) at least
every $\ell_l$ characters.

In Fig.\ \ref{Fig1}(b) we show our results using identical parameters to Fig.\
\ref{Fig1}(a), but using finite line lengths, $\ell_l = 20$, 30, 50, 70, 100,
and 200.  It can clearly be seen that the power law distribution is now
cut off by an exponential depending on the line length.  This behavior can
be understood in terms of the random walk formulation of the problem.  The
finite line length corresponds to an additional absorbing barrier for
the random walk, restricting $c_n \leq \ell_l\  \forall n$.  The corresponding
solution to the diffusion equation is
\begin{equation}
  f_f(x,t) = \sum_{j=1}^{\infty} C_j\,\sin(k_j\,x)\,\exp(-D\,k_j^2\,t)
\end{equation}
with long-time behavior dominated by $k_1$, where $k_j = \pi j/ \ell_l$.

\begin{figure}
  \vspace*{9.0cm}
  \includegraphics{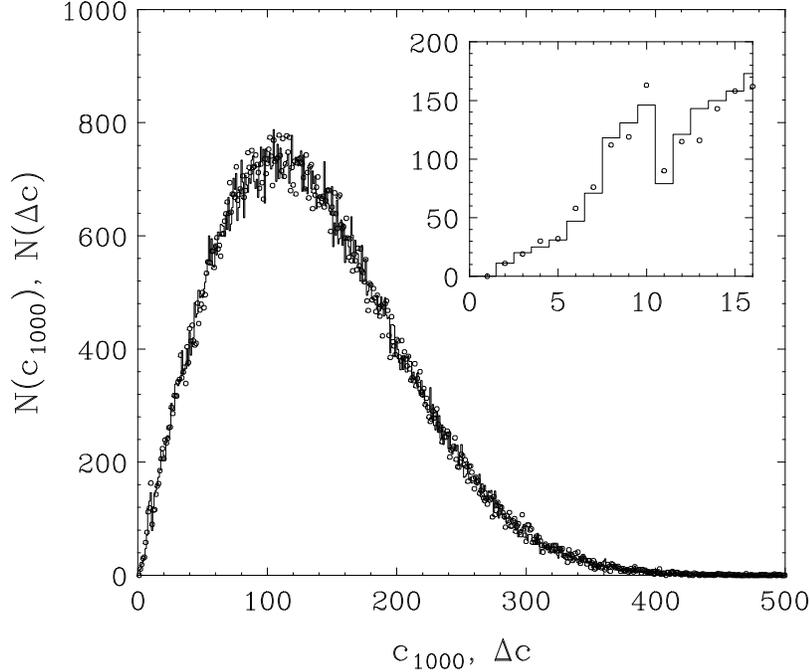}
  \caption[]{Histogram:  Number distribution, $N(\Delta c)$, of events with
      $\Delta c$ characters entered between catastrophic cascades, i.e.
      cascades which reached line 1000.  Circles:  Number distribution,
      $N(c_{1000})$, of events with total length of random walks, $c_{1000}$,
      for catastrophic cascades.  The inset is a magnified view of the region
      around the origin.  The parameters of this simulation are identical
      to the ones used for Fig.\ 1(a).}
  \label{Fig2}
\end{figure}

Of particular interest in studying models with `random' catastrophic events is
to investigate the limits of predictability of these events.  To do this we
record the number of characters, $\Delta c$, entered between catastrophic
cascades. Using the same parameters as for the calculations in Fig.\
\ref{Fig1}, we display in Fig.\ \ref{Fig2} (histogram) the number of events
as a function of $\Delta c$, $N(\Delta c)$.  It is obvious from this figure
that they follow a Wigner-distribution,
\begin{equation}
  N(\Delta c) \propto (\Delta c/\langle\Delta c\rangle)\,
               \exp(-\pi(\Delta c/\langle\Delta c\rangle)^2/4)\ .
\end{equation}
(As a side note we mention here that
we obtain essentially
identical behavior for finite line lengths, $\ell_l$, as long as
$\ell_l$ is large compared to the mean value of the
Wigner distribution, $\langle\Delta c\rangle$.)
Also displayed in this figure
(circles) is the distribution of the number of characters
pushed into
line 1000, $c_{1000}$.  This clearly follows the same functional form.
{}From our above considerations of the diffusion equation we see that
$N(c_{1000}) \rightarrow f(x,t\!=\!1000)$, and that therefore the mean
number of characters entered between catastrophic cascades is
\begin{equation}
  \langle\Delta c\rangle = \sqrt{\pi\, D\, n}
\end{equation}
We thus see that both distributions
displayed in Fig.\ \ref{Fig2} are governed by diffusion physics.

\begin{figure}
  \vspace*{8.7cm}
  \includegraphics{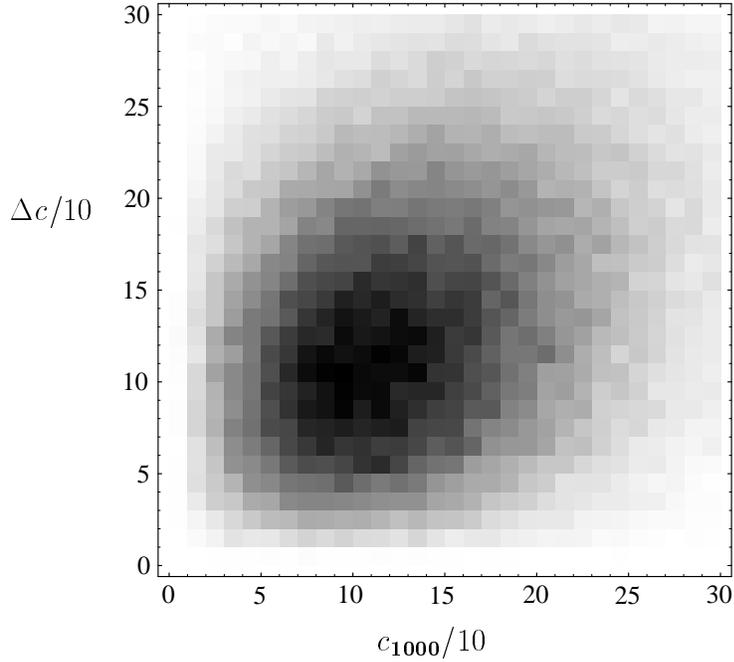}
  \caption[]{Contour plot of the number distribution, $N(c_{1000},\Delta c)$,
      of catastrophic cascades with random walks of total length $c_{1000}$
      and number of characters entered, $\Delta c$, before the next
      catastrophic cascade.  The grey level is proportional to
      $N(c_{1000},\Delta c)$, with black representing the maximum and white a
      value of 0.}
  \label{Fig3}
\end{figure}

Since $\Delta c$ represents the number of characters entered between
catastrophic cascades, and $c_{1000}$ is the number of characters removed
by a catastrophic cascade, sum rules require that $N(\Delta c)$ and
$N(c_{1000})$ have the same norm and mean.  The surprising aspect of
Fig.\ \ref {Fig2} is that both distributions are identical, even down to
the quadratic rise near the origin (see insert).  The connection with the
diffusion equation explains the Wigner-distribution form of $N(c_{1000})$,
but not for $N(\Delta c)$.

Despite the same functional shape, the above two distributions are not
tightly correlated.  Fig.\ \ref{Fig3} shows a density plot of the number
distribution
$N(c_{1000},\Delta c)$, where the grey level is proportional to the number of
counts in a given bin.
One sees only a weak enhancement
of this number distribution along the diagonal.
Here we plot the correlation between $c_{1000}$ and the time delay to the
next catastrophic cascade, but we obtain virtually identical results when
plotting the correlation between $c_{1000}$ and the time delay since the
previous catastrophic cascade.

The total number of trailing blanks summed over all lines up to a certain
maximum (here: 1000) changes by $-1$ each time a new character is entered.
Catastrophic cascades change the total number of blanks by $c_{1000}$.
Thus the total number of trailing blanks has exactly identical time dependence
as (up to a minus sign) the total mass of the sandpile measured in ref.\
\cite{RVR94}.

\begin{figure}
  \vspace*{10.2cm}
  \includegraphics{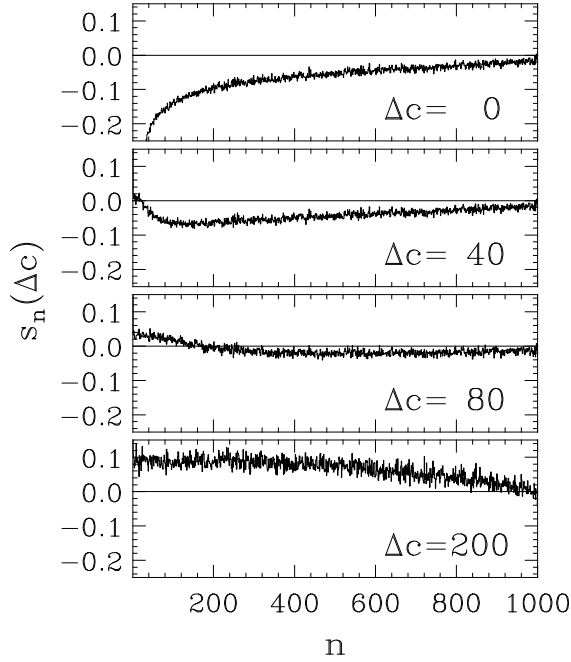}
  \caption[]{Average stress,
      $s_n(\Delta c)$, as a function of the line number, $n$,
      and the number of characters entered since the last catastrophic
      cascade, $\Delta c$.}
  \label{Fig4}
\end{figure}

In order to understand the trends of the buildup and release
we compute the average stress in each line,
\begin{equation}
  s_n(\Delta c) = \left\{\langle b_n\rangle - b_n\right\}_{\Delta c} ,
\end{equation}
as a function of the time delay since the last catastrophe.  (Here, the
notation $\{ ...\}_{\Delta c}$ indicates averaging over all events with
identical value of $\Delta c$.)
This is done in Fig.\
\ref{Fig4}.  We see from our data in Fig.\ \ref{Fig4}
that a catastrophic cascade typically inserts more extra blanks in the
early lines, thus reducing the stress in them below the average level.
This indicates that large positive steps early in the random walk are
correlated with catastrophic events.  However, the structure of the stress
as a function of the line number demonstrates the complex nature of the
evolution of the self-organized critical state.  This behavior cannot be
explained in terms of a simple random walk.

In conclusion, we have examined self-organized criticality in
line-wrap cascades in word processors.  We find that the distribution
of cascade lengths and cascade strengths can be accurately modeled
with the diffusion equation and compared to analytic forms.  We find
that the issue of predictability is complex.  Although the
distribution of times between large cascades is of a simple
Wigner-distribution form, stress develops in a rather complicated
style.  Even though the present system represents the simplest
non-trivial example of self ordered criticality, the model inspires a
wealth of questions, several of which we have addressed analytically
and many more which remain unresolved such as what is the optimum way
to predict the onset of large cascades.  By conquering this
easily-modeled example, insight may be reached regarding more complex
systems such as sand piles or more pertinent problems such as the
modeling of earth quakes.

It is also possible to extend the model to higher dimensionality or to
incorporate the addition of stress all through the paragraph rather
than only through the first line.  Work in these directions is
currently in progress.

This research was supported by NSF grant PHY-9403666, and
by an NSF Presidential
Faculty Fellow award (W.B.), NSF grant PHY-9253505.

\end{document}